\documentclass[]{mn2e}

\usepackage[dvips]{graphicx,color}
\usepackage{amssymb}
\usepackage{times}

\usepackage{amsmath}
\usepackage{pifont}
\usepackage{lscape}

\def\nms{\mathsurround=0pt}
\def\oversim#1#2{\lower 2pt\vbox{\baselineskip 
0pt \lineskip 1pt
    \ialign{$\nms#1\hfil##\hfil$\crcr#2\crcr\sim\crcr}}}
 \begin{document}

\title[Spectral Classification of O\,If*/WN stars]
{Spectral Classification of O2--3.5\,If*/WN5--7 stars}
\author[P. A. Crowther \& N. R. Walborn]{Paul A. Crowther$^{1}$\thanks{Paul.Crowther@shef.ac.uk},  
Nolan R. Walborn$^{2}$
\vspace{3mm} \\ 
$^{1}$Department of Physics and Astronomy, Hounsfield Road, University of 
Sheffield, Sheffield S3 7RH, UK\\
$^{2}$ Space Telescope Science Institute, 3700 San Martin Drive, 
Baltimore, MD 21218, USA\\
}
\date{\today}

\pagerange{\pageref{firstpage}--\pageref{lastpage}} \pubyear{2011}

\maketitle

\label{firstpage}

\begin{abstract} An updated classification scheme for transition 
O2--3.5\,If*/WN5--7  stars is presented, following recent revisions to the spectral 
classifications for O and WN stars. We 
propose that O2--3.5\,If*, O2--3.5\,If*/WN5--7 and WN5--7 stars may be
discriminated using the morphology of H$\beta$ to trace increasing wind 
density as follows: purely in absorption for O2--3.5\,If* stars in addition to the 
usual  diagnostics from Walborn et al.;   P Cygni for O2--3.5\,If*/WN5--7 stars; 
purely in emission for WN stars in   addition to  the usual diagnostics from Smith 
et al. We also discuss  approximate criteria to discriminate between these subtypes 
from near-IR  spectroscopy.  The physical and wind properties of such stars are 
qualitatively  discussed together with their  evolutionary significance. We suggest 
that the majority of O2--3.5\,If*/WN5--7 stars are  young, very massive  
hydrogen-burning stars, genuinely intermediate between  O2--3.5\,If* 
and WN5--7 subtypes, although a minority are apparently core helium-burning stars 
evolving blueward towards the classical WN sequence. Finally, we 
reassess  classifications for stars exhibiting lower ionization 
spectral features plus H$\beta$ emission.
\end{abstract}

\begin{keywords}
stars: early-type -- stars: evolution -- stars: fundamental parameters  -- stars: massive -- stars: Wolf-Rayet
\end{keywords}

\section{Introduction}

Historically, O3 stars have been considered to represent the highest mass 
main-sequence, i.e. core hydrogen-burning stars. 
However, it is now recognised that some hydrogen-rich, 
nitrogen sequence Wolf-Rayet stars may be main-sequence stars possessing still 
higher masses (de Koter et al. 1997; 
Schnurr et al. 2008; Smith \& Conti 2008; Crowther et al. 2010). 
Morphologically, it has long been recognised that there is a relatively smooth 
progression from O dwarfs through giants and supergiants to the WN sequence 
(Walborn 1971; Conti 1976; Crowther et al. 1995). Indeed, Walborn (1982a) 
introduced the hybrid O3\,If*/WN6 classification for Sanduleak --67$^{\circ}$ 22,
located in the Large Magellanic Cloud (LMC). Such stars, often 
referred to as
`hot' slash stars, possess intermediate spectral  characteristics between O3 supergiants (e.g. 
HD\,93129A) and WN6 stars (e.g. HD\,93162).  A second flavour of 
dichotomous spectrum, known as Ofpe/WN9 
or `cool' slash stars, was  introduced by Walborn (1982b) and Bohannan \& Walborn (1989), 
although alternative  WN9--11 subtypes are now in common usage for such stars (Smith et al. 1994).

Since the original study of O3\,If$^{\ast}$/WN6 stars by Walborn (1982a), the transition 
from 
photographic plates to digital detectors and increased samples have permitted the 
extension of the MK 
system to O2 (Walborn et al. 2002), while Smith et al. (1996) have added to the 
classification of WN stars. Indeed, the widespread availability of high quality 
spectroscopic datasets for O and WN stars -- such as the VLT-FLAMES Tarantula Survey (Evans 
et al. 2011) - allows us to reassess the hybrid Of/WN classification.

Here, we present a revised Of/WN classification scheme which takes
into account recent changes for Of and WN stars, based in part upon 
previously unpublished high-quality, blue-violet echelle spectrograms. Section 2 
describes archival and new Very Large Telescope (VLT) observations, while our scheme 
is described in Section 3. Section 4 provides an overview of Of/WN stars at
near-IR wavelengths, while a qualitative study of
such stars is presented in Section 5, together with a discussion of their
evolutionary significance. In Section 6 we briefly reassess spectral 
types of  Ofpe/WN9 plus related stars. Finally, a concise summary is 
presented in Section 7.

\begin{table*}
\begin{center}
\caption{Observing log of previously unpublished blue echelle spectroscopy used in 
this study}
\label{log}
\begin{tabular}{l@{\hspace{1.5mm}}r@{\hspace{1.5mm}}
c@{\hspace{1.5mm}}c@{\hspace{1.5mm}}l@{\hspace{2mm}}l@{\hspace{2mm}}c@{\hspace{2mm}}l@{\hspace{2mm}}l@{\hspace{2mm}}l}
\hline
Star & WR/   &Old&Ref.&VLT/ & Epoch&Sp. Coverage &Resolution& PI & Reference \\
     & BAT99-&Subtype  &         &Instrument &      &             &  km/\,s$^{-1}$        &    & \\

\hline
Sk --67$^{\circ}$ 22 &12 &O3\,If$^{\ast}$/WN6 &a  &UVES  &Dec 2004&3758-4983 & 8.7     &P.A. 
Crowther&Welty \& Crowther (2010) \\
TSWR 3  &93 &O3\,If$^{\ast}$/WN6 &b, d  &FLAMES&Dec 2008--Oct 2009 &3960--5114& 35--40
&C.J. Evans   &\#180: Evans et al. (2011)\\
Melnick 51&97 &O3\,If$^{\ast}$/WN7 &d  &FLAMES&Dec 2008--Oct 2009  &3960--5114& 35-40
&C.J. Evans   &\#457: Evans et al. (2011)\\
Melnick 39& 99&O3\,If$^{\ast}$/WN6 &d  &UVES  &Dec 2008-Jan 2010&4175-6200          
&7.5&C.J. Evans    &\#482: Evans et al. (2011)\\
Melnick 42&105&O3\,If$^{\ast}$/WN6 &d  &UVES  &Dec 2002&3758-4983          & 8.7 &P.A. Crowther&Welty \& Crowther 
(2010) \\
Melnick 30&113&O3\,If$^{\ast}$/WN6 &d  &FLAMES&Dec 2008--Oct 2009        
&3960-5114& 35-40&C.J. Evans   &\#542: Evans et al. (2011)\\
Melnick 35&114&O3\,If$^{\ast}$/WN6 &d  &UVES  &Dec 2008-Jan 2010&4175-6200          
&7.5&C.J. Evans    &\#545: Evans et al. (2011)\\
HD 38282  &118&WN6h       &c &UVES  &Nov 2003 &3300--6615& 4.5 &D. Welty   &Welty \& 
Crowther (2010) \\
\hline
\end{tabular}
\end{center}
(a) Walborn (1982a); (b) Testor \& Schild (1990); (c) Smith et al. (1996); (d) 
Walborn \& Blades (1997)
\end{table*}

\section{Observations}

Table~\ref{log} lists the previously unpublished echelle datasets used in this 
study, comprising objects within the LMC. All new datasets were 
obtained with the Very Large Telescope, using either the UV-Visual Echelle 
Spectrograph (UVES, D'Odorico et al. 2000) or the Medusa fibre-feed to the Giraffe
spectrograph of FLAMES (Pasquini et al. 2002).
%
%

UVES feeds both a blue (EEV CCD) and red (EEV CCD + MIT/LL CCD) arm, via various 
choices of dichroics and central wavelengths, with a small gap between red 
detectors. For the November 2002 and December 2004 service runs (70.D-0164, 
74.D-0109) two setups were used, together with a 1$''$ slit. The first used the blue 
and red arms of UVES, centred at 390/564nm, while the second used solely the red 
arm, centred at 520nm, providing complete spectral coverage between the far-blue and 
H$\alpha$. For the December 2003 run (72.C-0682) solely the 390/564nm setup was 
used, with a 0.7$''$ slit, producing a gap between blue and red arms from 
4500--4620\AA. Lower resolution spectroscopy from Crowther \& Smith (1997) obtained 
with the AAT/RGO spectrograph was used to fill this gap for HD\,38282. Details of 
data reduction are outlined in Welty \& Crowther (2010). 

UVES observations from the 
Tarantula survey (182.D-0222) also solely used the red arm, at 520nm, with a 1$''$ 
slit, which  omitted spectroscopy shortward of $\sim$4175\AA. In this instance, the 
crucial 
N\,{\sc iv} $\lambda$4058 region was obtained from archival Hubble Space Telescope 
(HST)/FOS spectrograph datasets from Massey \& Hunter (1998). Finally, multi-epoch 
Medusa/FLAMES datasets from the Tarantula survey were obtained with the 
(overlapping) LR02 and LR03 setups, providing complete blue-visual spectroscopy. 
Note that Melnick 30 and 51 were observed with Medusa/FLAMES and UVES, but the 
former are utilised here in view of the lack of complete blue coverage from the red 
arm of UVES. Details of data reduction are outlined in Evans et al. (2011). 
These 
datasets were complemented with archival high and intermediate dispersion 
observations of emission line stars obtained from a variety of sources, notably 
Crowther et al. (1995) and Walborn et al. (2002) for stars located within the Carina 
nebula. We note that all targets listed in Table~\ref{log} lie within the Large
Magellanic Cloud, in common with the majority of stars hitherto classified as 
O3\,If$^{\ast}$/WN6--7. Nevertheless, we will show that several Milky Way stars
also share these spectroscopic properties and discuss possible explanations for the
role of metallicity in Sect.~\ref{discussion}.

\section{H$\beta$ as a primary diagnostic for transition Of/WN stars}

In this section we present the motivation for an Of/WN sequence and 
our recommendations. Among early O-type stars, those with the highest wind 
densities exhibit He\,{\sc ii} $\lambda$4686 emission plus 
selective emission in N\,{\sc iii} $\lambda\lambda$4634--41 and N\,{\sc iv} 
$\lambda$4058, and are denoted Of$^{\ast}$ when the N\,{\sc iv} intensity is equal
to, or greater than, that of N\,{\sc iii}. Aside from these 
features, a conventional absorption line appearance is observed in 
the blue-violet spectra of such stars. Meanwhile, mid-to-late WN stars with 
relatively weak winds exhibit relatively weak, narrow He\,{\sc ii} 
$\lambda$4686 and N\,{\sc iv} $\lambda$4058 emission, 
with a P Cygni-type morphological appearance of the upper 
He\,{\sc ii} Pickering and/or H\,{\sc i} Balmer series. Nitrogen emission 
from N\,{\sc iii} $\lambda$4634--41 and N\,{\sc 
v} $\lambda\lambda$4603-20 is also common to such stars, typically 
corresponding to subtypes of WN5--7. Such stars have varying been labelled 
as WN-A (Hiltner \& Schild 1966; Walborn 1974), WN+abs (Smith 1968; Conti et al. 
1979), WN-w (Schmutz et al. 1989), WNha (Smith et al. 1996) and WNH stars 
(Conti \& Smith 2008).


Morphological similarities between Of* stars and such weak-lined, mid to 
late-type WN stars was first emphasised by Walborn (1971). Walborn (1982a) 
introduced Sk --67$^{\circ}$ 22 as a prototype for the O3\,If$^{\ast}$/WN6 subtype.
Following the identification of numerous early-type emission line stars in 30 
Doradus region of the LMC by Melnick (1985), several examples were so
classified (Walborn \& Blades 1997). 

The brightest stars of the central R136a ionizing cluster of 30 Doradus were also
initially classified as O3f/WN (Heap et al. 1994, de Koter et al. 1997) from UV 
spectroscopy, although WN4.5 or WN5 subtypes were preferred by Massey \& Hunter 
(1998) and Crowther \& Dessart (1998), respectively, from optical HST/FOS datasets.  
Still, in the  absence of  robust criteria,  individual stars have shifted between 
subclasses. For  example, Azzopardi  \& Breysacher (1979) initially classified 
their new LMC Wolf-Rayet star 
\#4 (AB4, Brey 58, BAT99-68) as WN5--6 while Smith et al. (1996) suggested
Of. Massey et al. (2000) supported WN5--6 for AB4 given its relatively strong 
He\,{\sc ii} $\lambda$4686 emission ($W_{\lambda}\sim$20\AA), although 
Massey et al. (2005) subsequently preferred O3\,If$^{\ast}$/WN6 on 
the basis of He\,{\sc ii} $\lambda$4200 absorption, while Schnurr et al. 
(2008a) assigned WN7h. Meanwhile, HD 93162 in the Carina Nebula has been 
traditionally described as a `weak-lined WN star' in spite of a lower He\,{\sc ii} 
$\lambda$4686 equivalent width ($W_{\lambda}\sim$15 \AA) than for AB4. Another 
example is Melnick  42,  initially classified as WN (Melnick 1982a) or O3\,If 
(Melnick 1985),  but subsequently reassigned to O3\,If$^{\ast}$/WN6 (Walborn et al. 1992).

\begin{table*}
\begin{center}
\caption{Horizontal criteria and standard (example) stars for O2--O4\,If 
(from Walborn et al. 2002, Sota et al. 2011), WN5--9 (revised from Smith et al. 1996) and
intermediate Of/WN stars, based on peak intensities of nitrogen diagnostics, 
N\,{\sc iii} $\lambda\lambda$4634--41, N\,{\sc iv} $\lambda$4058, 
N\,{\sc v} $\lambda\lambda$4603-20 plus H$\beta$. Differences with 
respect to Smith et al. (1996) are marked 
in bold.}
\label{sp_types}
\begin{tabular}{l@{\hspace{1.5mm}}l@{\hspace{1.5mm}}l@{\hspace{1.5mm}}l@{\hspace{1.5mm}}l@{\hspace{1.5mm}}l}
\hline
Subtype (H$\beta$ absorption) & O2\,If$^{\ast}$ & O3\,If$^{\ast}$ & O3.5\,If$^{\ast}$ & O4\,If & --  \\
\hline
Criteria  & N\,{\sc iv} em. $\gg$ N\,{\sc iii} em. &  N\,{\sc iv} em. $>$ N\,{\sc iii} em. & N\,{\sc iv} em. $\approx$ N\,{\sc iii} em. & N\,{\sc iv} em. $<$ N\,{\sc iii} 
em. & --\\
          & He\,{\sc i} absent &  He\,{\sc i} absent &  He\,{\sc i} absent & 
He\,{\sc i} weak & --\\
Standard stars & HD\,93129A & Cyg OB2-7, -22A & Pismis 24-1NE & HDE\,269698 & -- \\
&  &  &  & HD 190429A, Sk --67$^{\circ}$ 167 & -- \\
\hline
Subtype (H$\beta$ P Cygni) & O2\,If$^{\ast}$/WN5 & O2.5--3\,If$^{\ast}$/WN6 & O3.5\,If$^{\ast}$/WN7 & 
-- & --  \\
\hline
Criteria  & N\,{\sc iv} em. $\gg$ N\,{\sc iii} em. &  N\,{\sc iv} em. $>$ 
N\,{\sc iii} em. & N\,{\sc iv} em. $<$ N\,{\sc iii} em. & 
-- & --\\
          & N\,{\sc v} $\gtrsim $N\,{\sc iii} & N\,{\sc v} $<$ N\,{\sc 
iii} & N\,{\sc v} $\ll$ N\,{\sc iii} & -- & -- \\
Standard stars & Melnick 35 & HD 93162 (WR25) & Melnick 51  & 
-- & --\\
\hline
Subtype (H$\beta$ emission) & WN5 & WN6 & WN7 & WN8 & WN9 \\
\hline
Criteria & N\,{\sc v}/N\,{\sc iii} = 
{\bf 0.8} -- 2 & N\,{\sc v}/N\,{\sc iii} = 0.2 -- {\bf 0.8} & N\,{\sc iv}/N\,{\sc iii-v} = 
{\bf 0.3 - 0.8} & N\,{\sc iv}/N\,{\sc iii-v} 
$\leq$ {\bf 0.3} & N\,{\sc iv-v} absent, N\,{\sc iii} em. \\
         & N\,{\sc iv}/N\,{\sc iii-v} = {\bf 1-3} & N\,{\sc iv}/N{\sc iii-v} 
= {\bf 0.8 - 2} & N\,{\sc v}/N\,{\sc iii} $\leq$ {\bf 0.2} & 
N\,{\sc v}/N\,{\sc iii} $\leq$ {\bf 0.2} & P Cygni He\,{\sc i} \\
Standard stars & LS 2979 (WR49) & LS 3329 (WR67) & HD 151932 (WR78) & HD 96548 
(WR40) & NS4 (WR105) \\
\hline
\end{tabular}
\end{center}
\end{table*}

Since He\,{\sc ii} $\lambda$4686 emission, selective nitrogen emission 
plus intrinsic absorption components in the upper Pickering lines are 
common to some O2--3.5\,If$^{\ast}$ and WN5--7 stars, one needs to look elsewhere for a 
suitable diagnostic of intermediate O2--3.5\,If$^{\ast}$/WN5--7 stars, ideally in the 
conventional blue-visual range. We propose that a P Cygni morphology of 
H$\beta$ represents such a diagnostic, since this is uniquely in absorption 
for O stars (including Of* stars) and in emission for WN stars (though
see Sect.~\ref{oddities}). 
Various extensions to WN subtypes are in
common usage, for which h and a (or +abs) would be relevant to intermediate
Of/WN stars. We choose to omit these since they are redundant in such cases
(all contain hydrogen and intrinsic upper Pickering absorption lines).

Of course, 
two factors complicate the use of H$\beta$ as  a spectral diagnostic, 
namely nebular emission from an associated H\,{\sc ii} region, plus 
intrinsic absorption from a companion OB star. Consequently, Of/WN 
subtypes can only robustly be assigned on the basis of high dispersion, 
high S/N spectroscopy in which the nebular component and/or companion can 
be identified. 
Nevertheless, emission line strengths of other blue-violet features provide 
approximate subtypes (Sect.~\ref{zzz}) and high quality datasets are also 
required for reliable classification of the earliest O stars. 

\begin{figure*}
\begin{center}
\includegraphics[width=0.9\textwidth,clip]{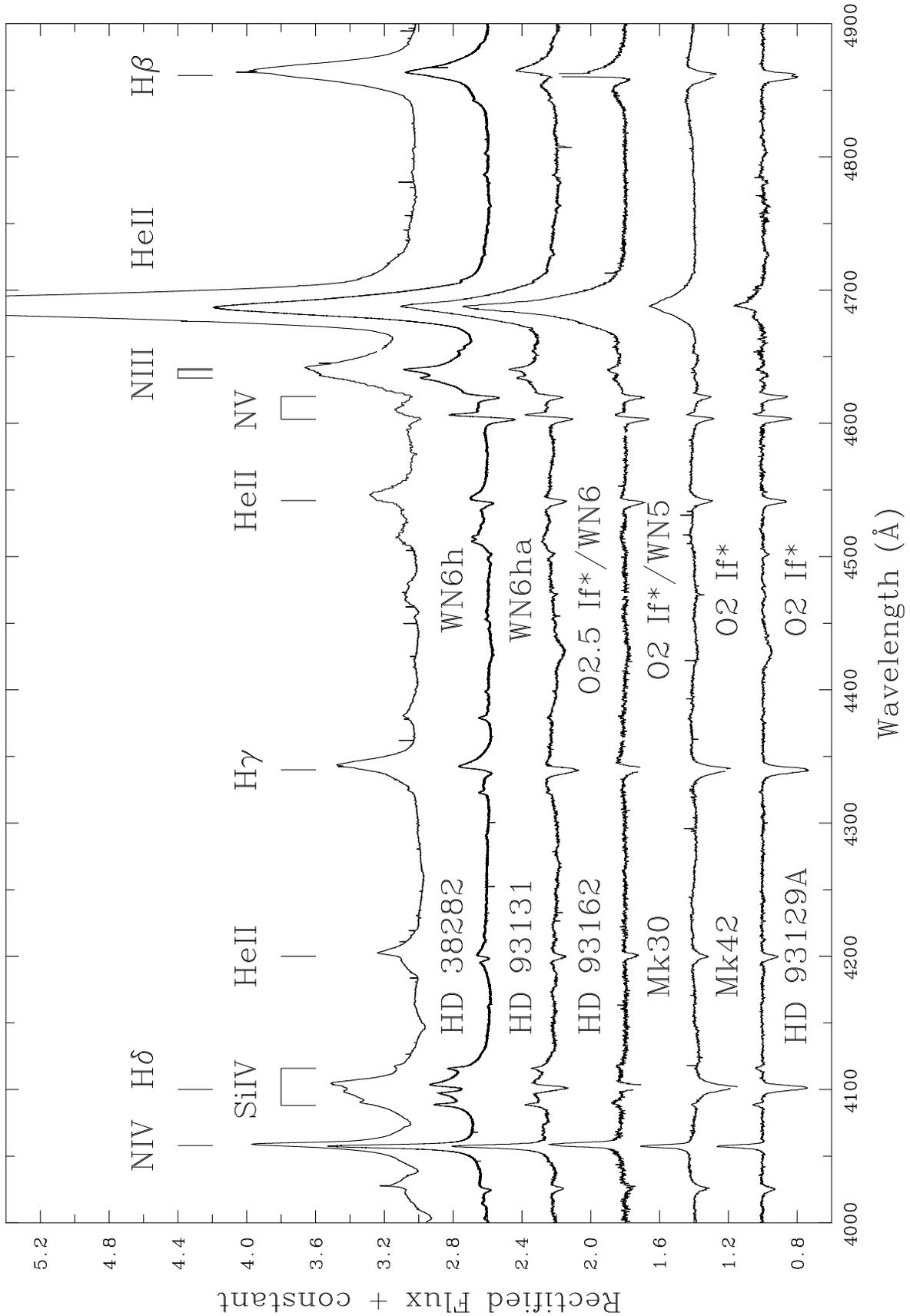}
\caption{Rectified, blue-violet spectrograms of stars spanning O2\,If$^{\ast}$ through WN6(h).
From H$\beta$, Melnick 42 is newly classified O2\,If$^{\ast}$ while HD 93162 is revised to 
O2.5\,If$^{\ast}$/WN6h.
Stars are uniformly offset by 0.4 continuum units for clarity.}
\label{ofwn_lab}
\end{center}
\end{figure*}

Our use of H$\beta$ to discriminate between subclasses has a relatively minor 
effect on existing spectral subtypes, although a few stars shift between 
categories. In addition it is  necessary to adjust the division between WN5 and WN6 
subtypes with respect to Smith et al. (1996), since the parallel Of and WN 
sequences both involve the N\,{\sc iv}/N\,{\sc iii} ratio. For hybrid Of/WN 
subtypes, universally exhibiting N\,{\sc iv} $\lambda$4058 emission, the range in 
ionization spans O2--O4\,If and WN5--8, with a relatively monotonic 
sequence from O2\,If$^{\ast}$/WN5 to O3.5\,If$^{\ast}$/WN7. For completeness, our updated 
criteria are set out in Table~\ref{sp_types}, for which horizontal criteria reflect 
changes of ionization (decreasing stellar temperatures from left to right)  and vertical criteria 
indicate changes in wind density (increasing from top to bottom).

\subsection{Morphological sequence from O2 to WN5--6}

In Figure~\ref{ofwn_lab} we present representative examples of the highest 
ionization stars  spanning the morphological sequence O2\,If$^{\ast}$, O2\,If$^{\ast}$/WN5--6, 
WN5--6. With 
respect to current classification schemes, it was necessary for Melnick 42 
(Melnick 1985, Walborn \& Blades 1997) to be reassigned to a O2\,If$^{\ast}$ classification 
since its overall morphology more closely resembles HD 93129A (Walborn et 
al. 2002) than Melnick 30 (Walborn \& Blades 1997) which is newly revised from 
O3\,If/WN6 to  O2\,If$^{\ast}$/WN5. The Figure illustrates the significance of the H$\beta$ 
morphology, with evidence for P Cygni profiles in H$\gamma$ and He\,{\sc ii} 
$\lambda$4542 in Melnick 30. 

Similarly, we reassign HD 93162 from WN6ha (Smith et al. 1996)
to an intermediate O2.5\,If$^{\ast}$/WN6 classification on the 
basis of
P Cygni H$\beta$, in common with Melnick 30, rather than emission as is the case of 
HD 93131 (WN6ha). Conti \& Bohannan (1989) have previously highlighted its
intermediate morphological appearance by suggesting WN6/O4f for  HD 93162.
Evans et al. (2006) introduced the O2.5 subclass for  N11-026 
(O2.5\,III(f*)) since its appearance lay intermediate between O2 and O3 
standards from Walborn et al. (2002). 

\begin{figure*}
\begin{center}
\includegraphics[width=0.9\textwidth,clip]{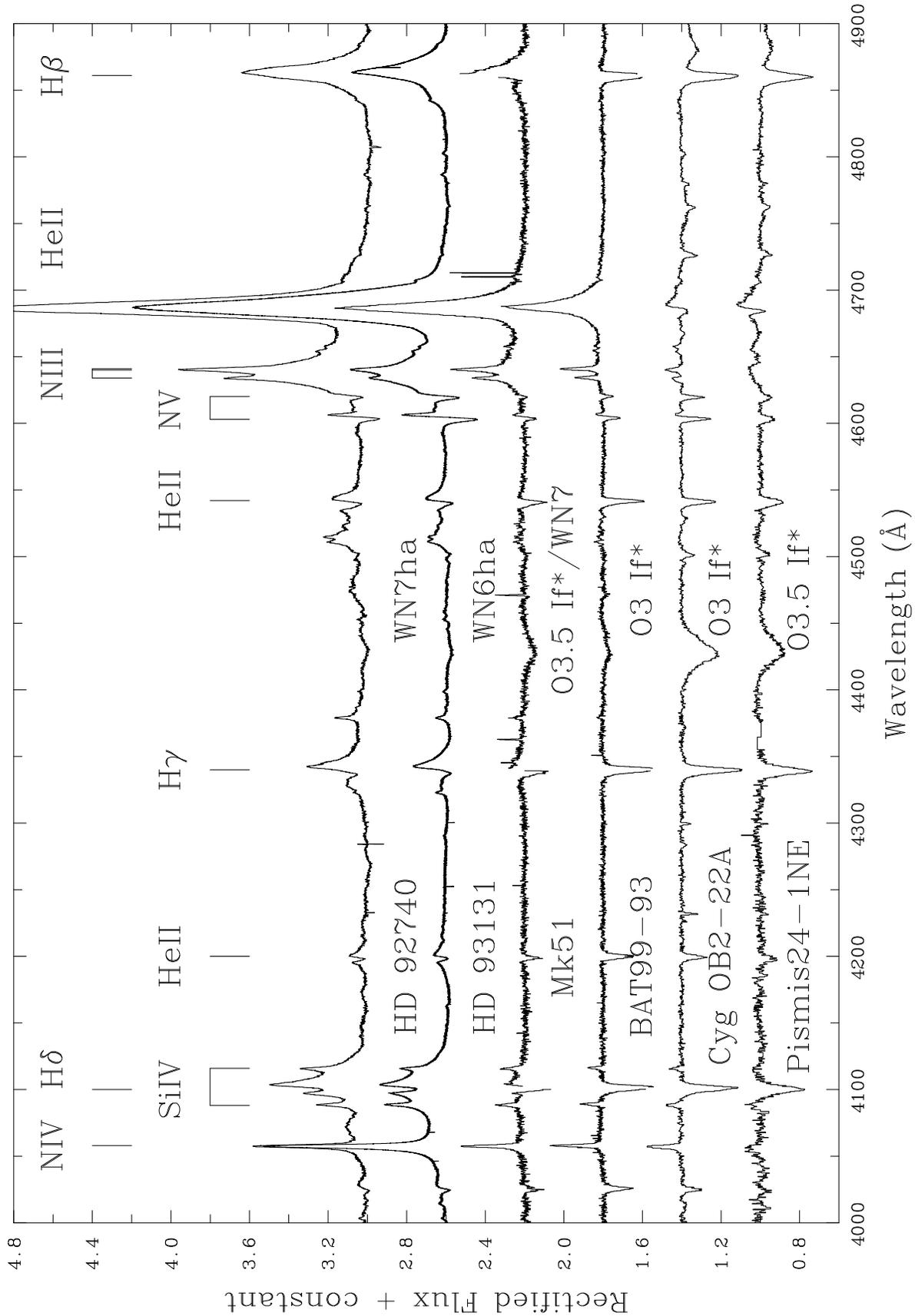}
\caption{Rectified, blue-violet spectrograms of stars spanning O3\,If$^{\ast}$ through WN6--7. 
From H$\beta$, BAT99-93 (TSWR3, Brey 74a) is 
newly classified O3\,If$^{\ast}$ while Melnick 51 is refined to O3.5\,If$^{\ast}$/WN7.
Stars are uniformly offset by 0.4 continuum units for clarity.}
\label{ofwn_mid_lab}
\end{center}
\end{figure*}

\subsection{Morphological sequence from O3--3.5 to WN6--7}

In Figure~\ref{ofwn_mid_lab} we present representative examples of stars of 
slightly lower ionization, 
spanning the morphological sequence O3--3.5\,If$^{\ast}$, O3--3.5\,If$^{\ast}$/WN6--7, 
WN6--7, including O supergiants from Walborn et al. (2002) and 
Ma\'{i}z Apell\'{a}niz et al. (2007). With respect to existing classification 
schemes, it was necessary for BAT99-93 (Testor \&  Schild 1990, Walborn  \& Blades 
1997) to be classified as O3\,If$^{\ast}$  since  H$\beta$ is in absorption, in spite of 
its prominent He\,{\sc ii} $\lambda$4686 emission. Meanwhile, O3.5\,If$^{\ast}$/WN7 is 
preferred for Melnick 51  since the morphology of this star is intermediate between 
BAT99-93 and HD 92740, even though the P Cygni nature is H$\beta$ is less 
definitive than other stars. 

It could be argued that Melnick 51 should be assigned O3.5\,If$^{\ast}$/WN6.5 
since N\,{\sc iv} $\lambda$4058 $\sim$ N\,{\sc  iii} 
$\lambda\lambda$4634-41, i.e. intermediate between N\,{\sc iv} 
$\lambda$4058 $>$ N\,{\sc  iii} $\lambda\lambda$4634-41 for 
non-transition WN6 stars and N\,{\sc iv} $\lambda$4058 $<$ N\,{\sc  iii} 
$\lambda\lambda$4634-41 at a subtype of WN7. For the moment we prefer WN7 
for Mk 51 on  the basis of N\,{\sc  v} $\lambda\lambda$4603-20 $\ll$ 
N\,{\sc iii} $\lambda\lambda$4634-41, in common with non-transition WN7 
stars. However, should other examples of similar transition stars
be confirmed (Mk 37a is a candidate), we may  
reconsider the use of WN6.5 for intermediate narrow-lined types (N\,{\sc 
iii-v} lines are severely blended for broad-lined stars).

\begin{figure*}
\begin{center}
\includegraphics[width=0.9\textwidth,clip]{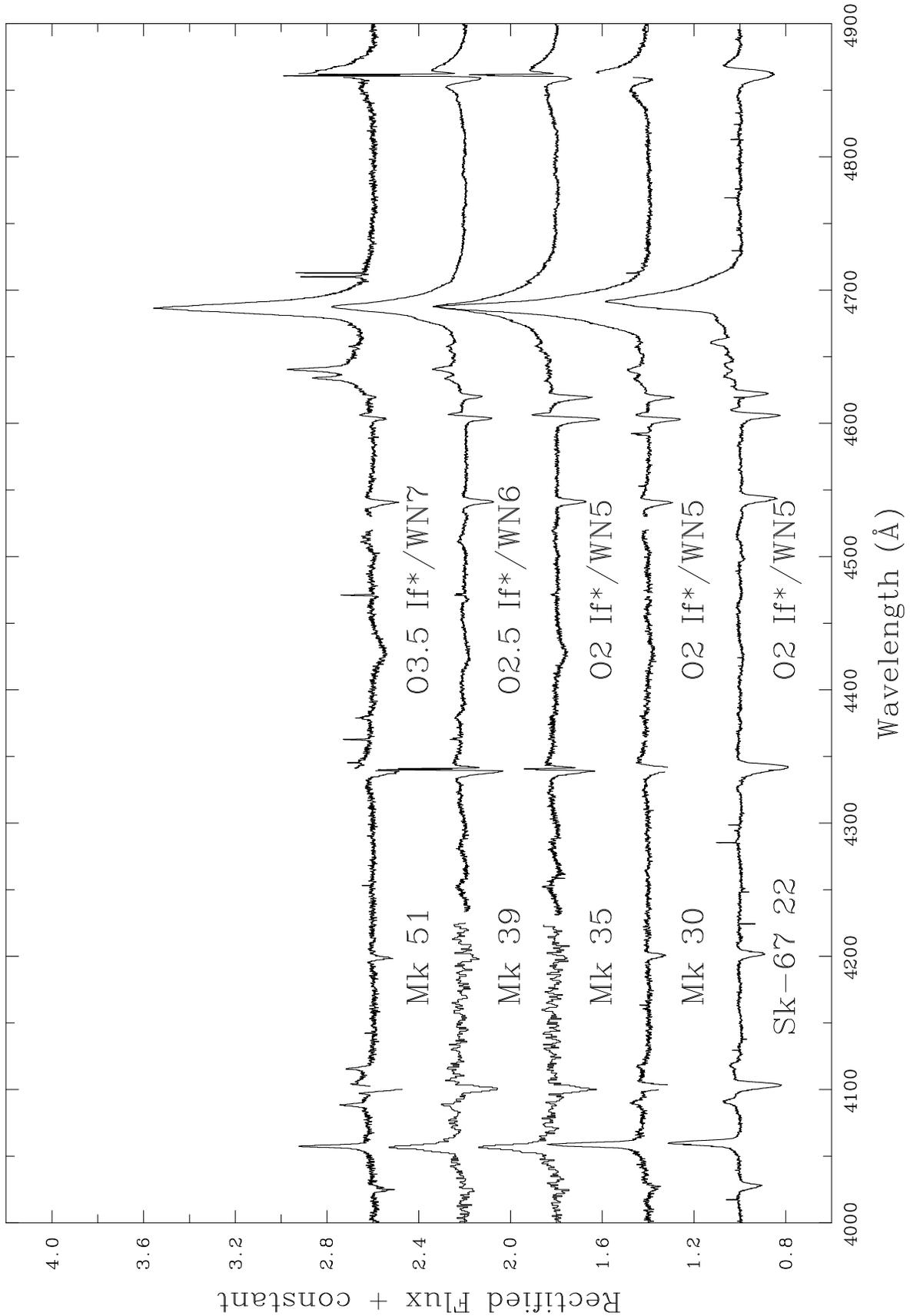}
\caption{Rectified, blue-violet spectrograms of transition Of/WN stars.
Although H$\beta$ is P Cygni for all cases, there is a continuum from the least 
(Sk --67$^{\circ}$ 22) to the most extreme (Melnick 51). Stars are uniformly offset by 0.4 continuum units for 
clarity.}
\label{ofwn_montage_lab}
\end{center}
\end{figure*}



\subsection{Morphological sequence from O2\,If$^{\ast}$/WN5 to O3.5\,If$^{\ast}$/WN7}

Figure~\ref{ofwn_montage_lab} presents a montage of 
Of/WN stars for which we possess high quality blue-violet 
spectroscopy. In two cases -- Melnick 35 and 39 -- the N\,{\sc iv} 
$\lambda$4058 region was not included in UVES datasets from the VLT-FLAMES
Tarantula Survey (Evans et al. 2011), 
so  lower resolution  archival HST/FOS spectroscopy of this 
region has been included (Massey \& Hunter  1998).

This figure illustrates the range in ionization balance sampled by 
transition Of/WN stars, given the requirement that N\,{\sc iv} 
$\lambda$4058 emission is observed in all instances. By definition, P 
Cygni profiles are observed for H$\beta$, although there is a continuum 
from Sk --67$^{\circ}$ 22 (least extreme)  to Melnick 51 (most extreme). 
Indeed, Sk --67$^{\circ}$ 22 is the only example from this sample in which 
H$\gamma$ is purely in absorption, rather than a P Cygni profile.

\subsection{Subtype boundaries from optical spectroscopy}\label{zzz}

In addition to the new high dispersion observations set out in 
Table~\ref{log}, we have reassessed spectral types for other early O 
supergiants and weak-lined WN stars in the Milky Way and LMC based upon lower 
resolution spectroscopy.  Spectral  types from the literature, plus our new 
revisions where necessary, are provided in Table~\ref{catalog}.

\begin{figure*}
\begin{center}
\includegraphics[width=0.9\textwidth,clip]{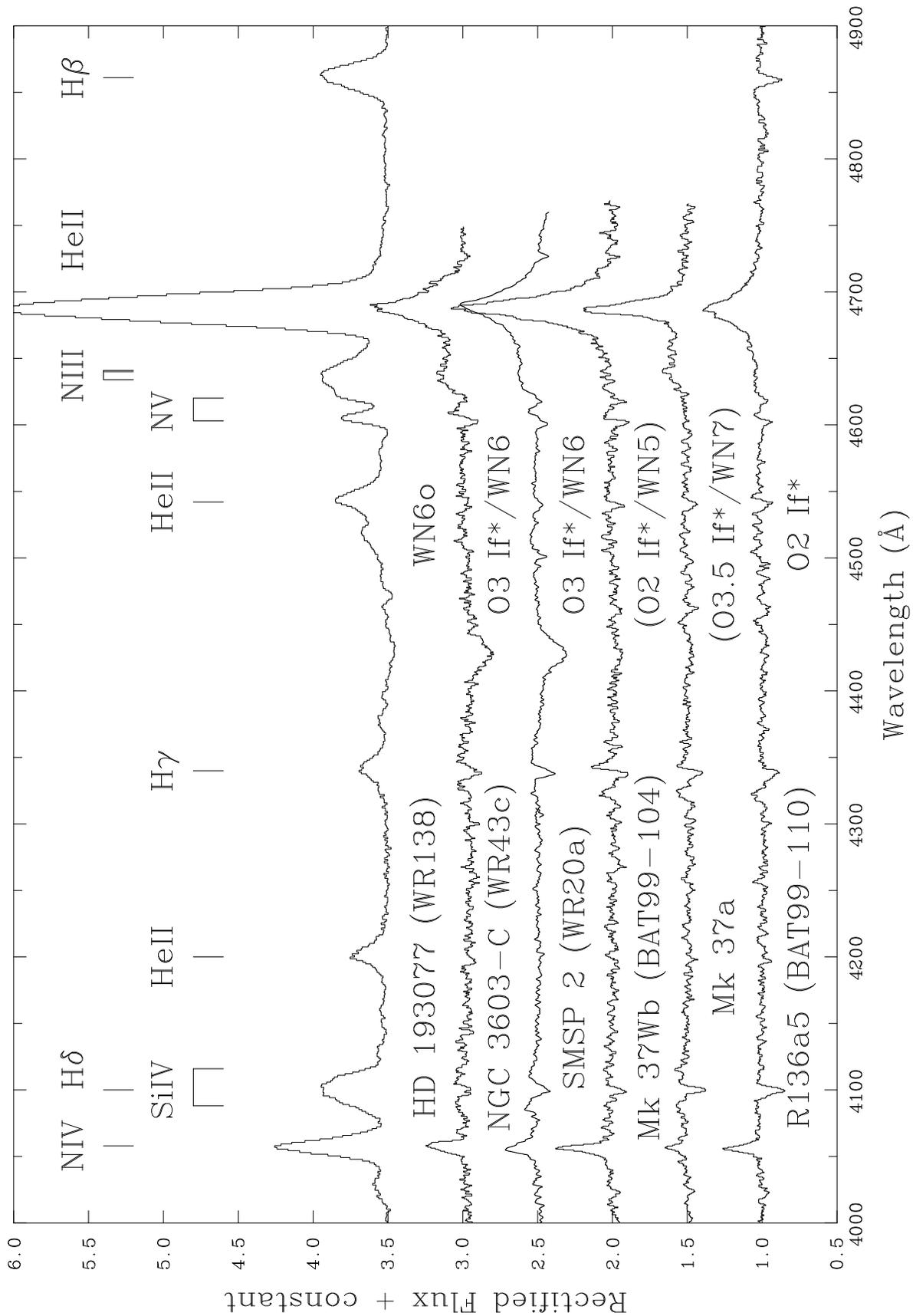}
\caption{Rectified, blue-violet spectrograms of early-type, emission line stars, 
for which revised spectral  classifications are obtained (provisional subtypes are indicated in parentheses 
for objects lacking published H$\beta$ spectroscopy). Stars are uniformly offset by 0.5 continuum units for clarity.}
\label{newbies}
\end{center}
\end{figure*}

In Figure~\ref{newbies} we present blue-violet spectrograms for stars that 
we have revised with respect to recent literature values. In the two
cases for which datasets include H$\beta$, we can comfortably assign 
O2\,If$^{\ast}$ to R136a5 and WN6o to HD 193077 (WR138). 
Although our spectroscopy does not extend to H$\beta$ for NGC 3603-C (WR43c),
figure~3 from Melena et al. (2008) indicates an emission morphology with weak 
P Cygni absorption. NGC 3603-C has  been classified as WN6+abs and WN6ha by Drissen et 
al. (1995) and Schnurr et al. (2008b), respectively, although the 
former authors noted its striking similarity to HD\,93162. We assign a 
slightly later subtype of O3\,If$^{\ast}$/WN6 for NGC 3603-C given its lower 
N\,{\sc iv}/N\,{\sc iii} ratio than HD\,93162. 

As for NGC 3603-C, we do not possess H$\beta$ spectroscopy for SMSP2 (WR20a). Fortunately,
H$\beta$ emission is clearly observed in figure~7 of 
Shara et al. (1991), who assigned
a WN7 subtype for WR20a. Rauw et al. (2004) discovered the binary nature 
of WR20a and obtained
spectral types of O3\,If$^{\ast}$/WN6 + O3\,If$^{\ast}$/WN6. This was 
subsequently revised
to WN6ha + WN6ha by Rauw et al. (2005) since the emission equivalent width 
of   He\,{\sc ii} $\lambda$4686 narrowly exceeded  12\AA, representing a 
boundary
proposed by Crowther \& Dessart (1998) to accommodate HD 93162 within the WN sequence. Of 
course, since we have newly reassigned HD 93162 from WN6ha to O2.5\,If$^{\ast}$/WN6 star, this 
criterion no longer applies (see above). For WR20a we find an identical 
N\,{\sc iv}/N\,{\sc iii} ratio to  WR43c so assign O3\,If$^{\ast}$/WN6 for each of the 
components in this system. 

\begin{figure*} \begin{center} 
\includegraphics[width=0.9\textwidth,clip]{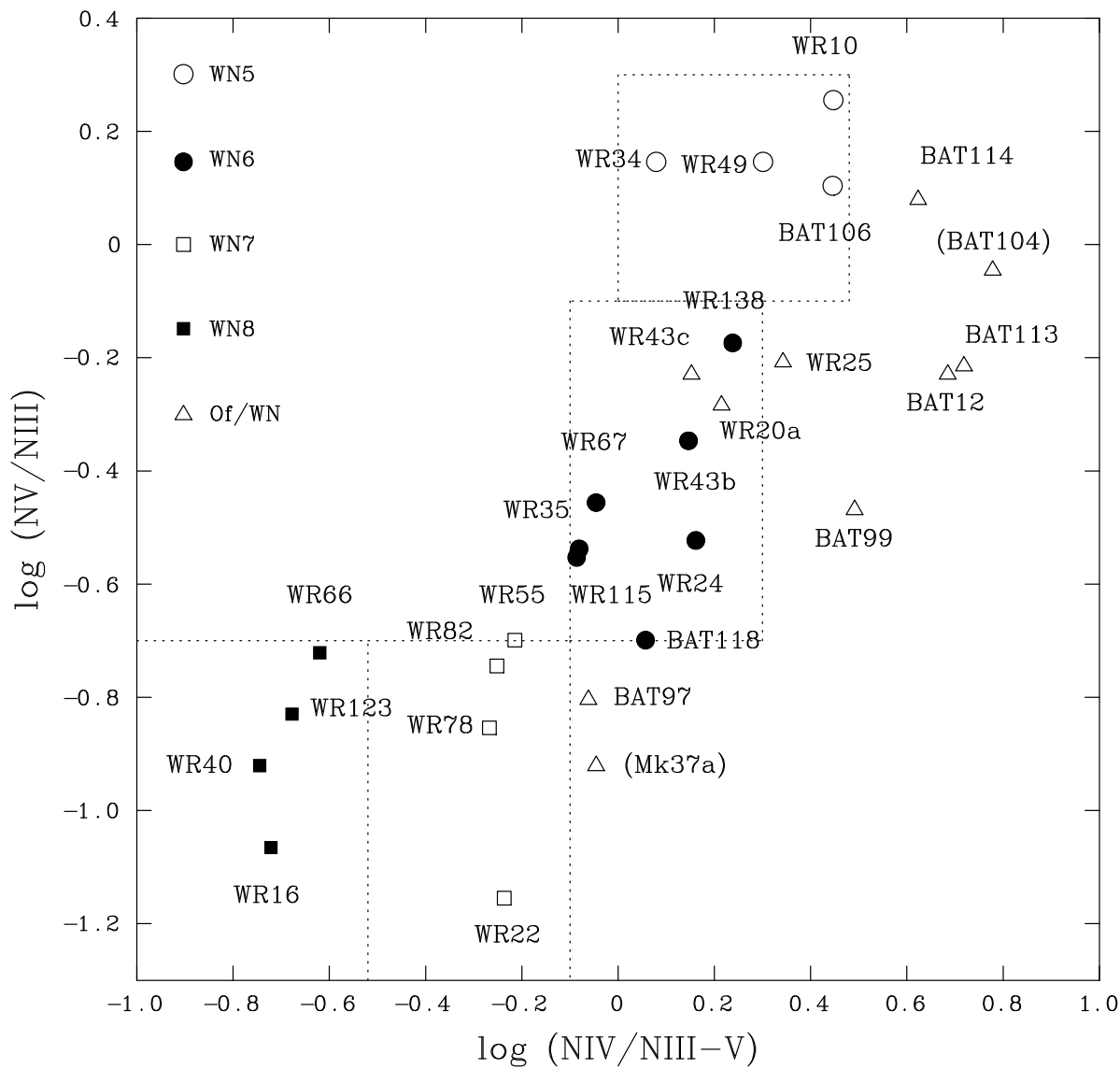} \caption{Nitrogen 
line ratios of selected O2--3.5\,If$^{\ast}$/WN5--7 and WN5--8 stars, based upon 
our revised WN subtype boundaries (dotted lines). Preliminary spectral types
are indicated in parentheses.} \label{plot_log} 
\end{center} \end{figure*}

In two further cases, Mk 37a and Mk 37Wb, published spectroscopy does not extend to H$\beta$.
Their subtypes are therefore provisional, although we defer their discussion until we
have considered their nitrogen line ratios and He\,{\sc ii} $\lambda$4686
line strengths in the context of other early-type emission line stars.
We illustrate the nitrogen line ratios of selected O2--3.5\,If$^{\ast}$/WN5--7 and 
WN5--8 stars in Figure~\ref{plot_log}, including subtype boundaries set 
out in Table~\ref{sp_types}. Of/WN stars tend to possess higher ratios of 
N\,{\sc iv} $\lambda$4058/N\,{\sc iii-v} $\lambda\lambda$4603--41 than WN 
stars, although  O2\,If$^{\ast}$/WN5 stars exhibit reduced ratios of N\,{\sc v} 
$\lambda\lambda$4603--20/N\,{\sc iii} $\lambda\lambda$4634--41 
 with respect to WN5 stars.

\begin{figure*} \begin{center} 
\includegraphics[width=0.9\textwidth,clip]{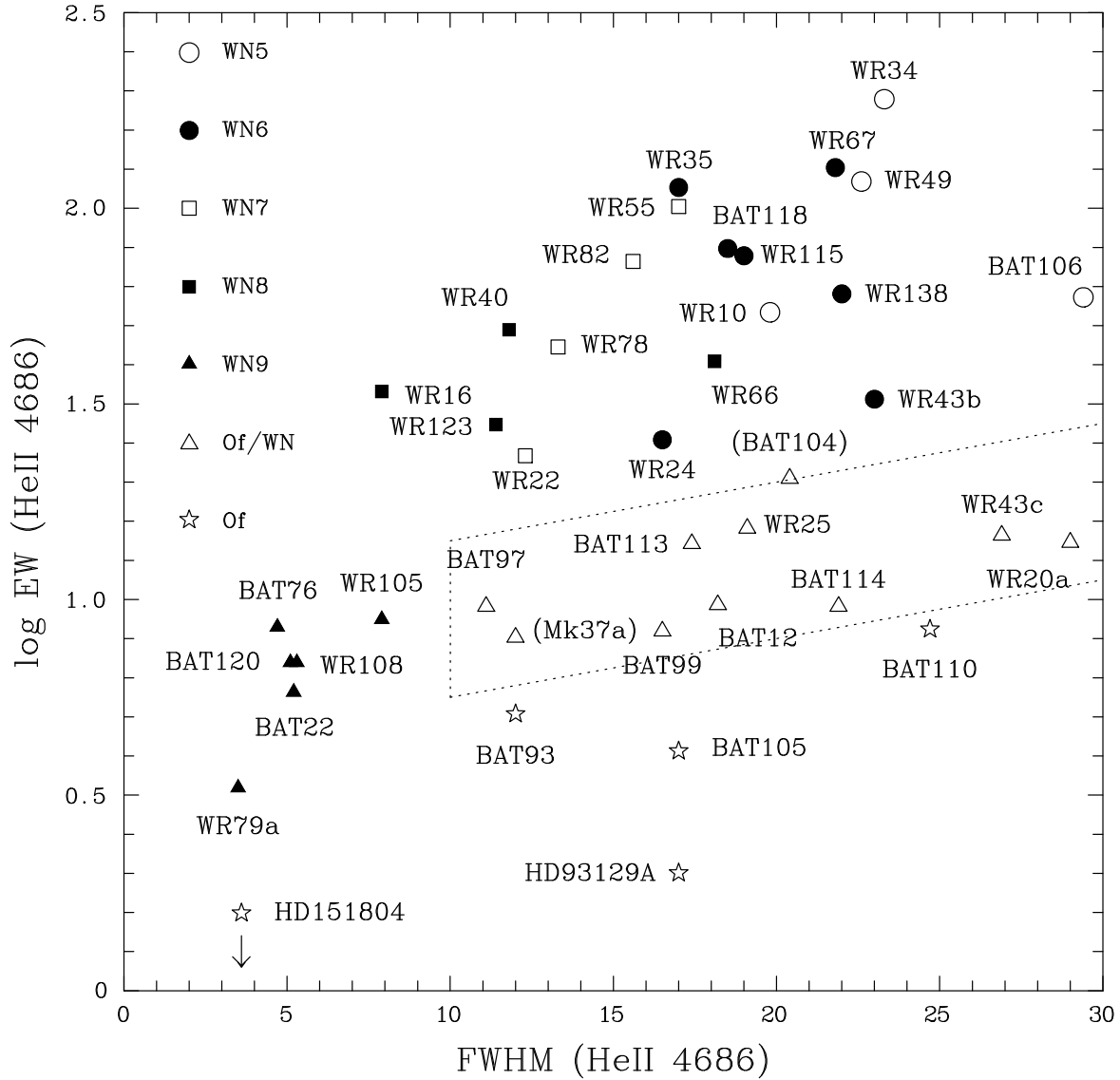} \caption{Comparison 
between He\,{\sc ii} $\lambda$4686 line strength (equivalent width in \AA) 
and full width at half maximum (in \AA) for selected WN5--9 stars, 
O2--3.5\,If$^{\ast}$/WN5--7 stars and O2--3.5\,If$^{\ast}$ stars. An approximate
boundary for Of/WN stars is indicated (dotted lines). Preliminary spectral 
types are indicated in parentheses.} 
\label{4686} 
\end{center} \end{figure*}

Since He\,{\sc ii} $\lambda$4686 is the most prominent emission line in 
the blue-visual spectrum of O-type supergiants and Wolf-Rayet stars, we 
now assess whether this line alone allows a suitable discriminator 
between O\,If$^{\ast}$, O\,If$^{\ast}$/WN5--7 and WN5--7 stars. Figure~\ref{4686} compares 
the line strength and line width of $\lambda$4686 for various 
emission-line stars, to which we have added several examples of WN8--9 stars. 
O2--3.5\,If$^{\ast}$/WN5--7 stars indeed possess properties intermediate between 
O2--3.5\,If$^{\ast}$ stars and WN5--7 stars, with 8\AA\ $\leq$ EW(He\,{\sc ii} 
$\lambda$4686) $\leq$ 20\AA, and 10\AA\ $\leq$ FWHM(He\,{\sc ii} 
$\lambda$4686) $\leq$30\AA. 

It is apparent that the emission equivalent width of 
He\,{\sc ii} $\lambda$4686 alone does not permit an 
unambiguous subtype. R136a5 (BAT99-110, O2\,If$^{\ast}$) possesses a similar line 
strength to Mk\,39 (BAT99-99, O2\,If$^{\ast}$/WN5). We indicate
approximate (inclined) boundaries for Of/WN subtypes in Fig.~\ref{4686}.
To illustrate difficulties close to subtype boundaries we 
reconsider the spectral type for AB4 (Brey 58, BAT99-68) which has 
fluctuated between Of/WN and WN subtypes in the literature.  

We have been provided with a digital version of the blue spectrum for AB4  
presented by Massey et al. (2005) which extends beyond H$\beta$.  Its 
H$\beta$ morphology very closely resembles Melnick 51  (O3.5\,If$^{\ast}$/WN7), 
although He\,{\sc ii} $\lambda$4686 is stronger in emission for AB4
and H$\gamma$  is a more developed P Cygni profile. Since we assign 
O3.5\,If$^{\ast}$/WN7 for  Melnick 51 and identical subtype is 
preferred for BAT99-68, although  its He\,{\sc ii} $\lambda$4686 emission strength and 
width sits atop the boundary between Of/WN and WN stars presented in Fig.~\ref{4686}. 
On balance we favour the intermediate Of/WN subtype proposed by 
Massey et al. (2005), albeit at a somewhat later subclass.

\begin{enumerate}

\item {\bf Mk\,37a}
We provide a minor revision to the spectral type of Mk\,37a from O4\, If
(Massey \& Hunter 1998) to O3.5\,If$^{\ast}$ following the updated classification 
scheme of Walborn et al. (2002) for early O stars, since N\,{\sc iv} 
$\lambda$4058 $\sim$ N\,{\sc iii} $\lambda\lambda$4634--41. The He\,{\sc ii}
$\lambda$4686 line strength and width for this star are similar to Mk 51 in 
Figure~\ref{4686}, so we favour O3.5\,If$^{\ast}$/WN7, although 
H$\beta$ spectroscopy is strictly required for an unambiguous classification.

\item {\bf Mk 37Wb (BAT99-104)} Finally, we reassess the spectral type of 
O3\,If$^{\ast}$/WN6 for Mk 37Wb by Massey \& Hunter (1998). Its overall 
blue-violet morphology matches that of confirmed Of/WN stars (e.g. Mk 35), 
although it is located close to the boundary between Of/WN and WN stars in 
Fig.~\ref{4686}, since $W_{\lambda}$(He\,{\sc ii} $\lambda$4686) 
$\sim$20\AA and FWHM(He\,{\sc ii} $\lambda$4686)$\sim$20\AA. For the 
moment we suggest a minor revision to its spectral type from 
O3\,If$^{\ast}$/WN6 to O2\,If$^{\ast}$/WN5, since N\,{\sc iv} 
$\lambda$4058 $\gg$ N\,{\sc iii} $\lambda\lambda$4634--41, although we are 
unable to provide a definitive subtype in the absence of H$\beta$ 
spectroscopy. \end{enumerate}

\begin{table*}
\begin{center}
\caption{Catalogue of selected Milky Way and LMC O2--3.5\,If$^{\ast}$  supergiants, O2--3.5\,If$^{\ast}$/WN5--7 
and weak-lined WN5--7 stars,
including revisions to literature spectral types, sorted by absolute K-band magnitude. For stars lacking H$\beta$ spectroscopy,
spectral types are provisional and so are shown in parenthesis. Near-IR photometry
is from 2MASS except where noted, while distances are obtained as follows: LMC (49 kpc, Gibson 2000), NGC 3603 (7.6 kpc,
Melena et al. 2008), Westerlund 2 (7.9 kpc, Rauw et al. 2005), Carina Nebula (2.3 kpc, Davidson \& Humphreys 1997), Pismis 24
(2.5 kpc, Massey et al. 2001), Cyg OB2 (2 kpc, Massey \& Thompson 1991), Cyg OB1 (2 kpc, Humphreys 1978).}
\label{catalog}
\begin{tabular}{l@{\hspace{1.5mm}}l@{\hspace{1.5mm}}l@{\hspace{1.5mm}}l@{\hspace{1.5mm}}l@{\hspace{1.5mm}}l@{\hspace{1.5mm}}
l@{\hspace{1.5mm}}l@{\hspace{1.5mm}}l@{\hspace{1.5mm}}l@{\hspace{1.5mm}}l@{\hspace{1.5mm}}l}
\hline
Star & Alias & Old &  Ref & New & $m_{\rm K}$ & $A_{\rm K}$ & Ref & DM & Note & $M_{\rm K}$ & $M_{\rm Bol}$ \\
     &       & Subtype & & Subtype & mag & mag && mag & & mag & mag \\
\hline
HD 38282 & BAT99-118 & WN6h & a & & 10.6 & 0.2 & & 18.45 & LMC & --7.9 & --12.0  \\ 
AB4 & BAT99-68 & O3\,If$^{\ast}$/WN6, WN7h & b,c & O3.5\,If$^{\ast}$/WN7 & 11.2: & 0.7: & & 18.45 & LMC & --7.9: 
& --11.6: \\ 
R136a1 & BAT99-108 & WN5h & d & & 11.1 & 0.2 & m & 18.45 & LMC & --7.6 & --12.5 \\
NGC 3603-A1 & WR43a & WN6ha+WN6ha & d, e & & 7.4 & 0.6 & m & 14.4 & NGC 3603 & --7.6 & --11.7\\
SMSP2 & WR20a & WN7, WN6ha+WN6ha & f, g & O3\,If$^{\ast}$/WN6 + O3\,If$^{\ast}$/WN6 & 7.6 & 0.7 & & 14.5 
& Wd 2 & --7.6 & --12.0 \\
NGC 3603-B & WR43b  & WN6ha & d, e & & 7.4 & 0.6 & m & 14.4 & NGC 3603 & --7.5 & --11.6  \\
R136c  & BAT99-112 & WN5h & d & & 11.3 & 0.3 & m & 18.45 & LMC & --7.4 & --12.3\\
R136a2 & BAT99-109 & WN5h & d & & 11.4 & 0.2 & m & 18.45 & LMC & --7.3 & --12.2  \\
Mk 34 & BAT99-116 & WN5h & d & & 11.7 & 0.3 & m & 18.45 & LMC & --7.0 & --11.9 \\
R136a3 & BAT99-106 & WN5h & d & & 11.7 & 0.2 & m & 18.45 & LMC & --6.9 & --11.8  \\
HDE 319718NE & Pismis 24-1NE & O3.5\,If$^{\ast}$  & h, i   &  & 5.9 & 0.7 & & 12.0 & Pismis 24 & --6.8 & --11.4 \\
NGC 3603-C & WR43c & WN6ha & d, e & O3\,If$^{\ast}$/WN6 & 8.3 & 0.6 & m &  14.4 &NGC 3603 &  --6.7 & --11.1 \\
HD 92740 & WR22 & WN7ha & a & & 5.4 & 0.1 && 11.8 & Carina & --6.5 & --10.2 \\
Mk 39  & BAT99-99  & O3\,If$^{\ast}$/WN6 & j, k & O2.5\,If$^{\ast}$/WN6 & 12.1 & 0.2 & m & 18.45 &LMC & --6.5 & --11.7  \\
HD 93162 & WR25 & WN6ha & a & O2.5\,If$^{\ast}$/WN6 & 5.7 & 0.3 & n & 11.8 & Tr 16 & --6.4 & --10.8 \\ 
Mk 42  & BAT99-105 & O3\,If$^{\ast}$/WN6 & j, k & O2\,If$^{\ast}$ & 12.2 & 0.2 & m & 18.45 & LMC & --6.4 & --11.6 \\
Mk 37a &           & O4\,If & k     & (O3.5\,If$^{\ast}$/WN7) & 12.4 & 0.2 & m & 18.45 & LMC & --6.3 & --10.9  \\
HD 93129A &        & O2\,If$^{\ast}$ & h & & 6.0  & 0.4  &&  11.8 & Tr 14 &  --6.2  & --11.4 \\ 
HD 93131 & WR24 & WN6ha & a &  & 5.8 & 0.1 & n & 11.8 & Col 228 & --6.1 & --10.2  \\ 
R136a5 & BAT99-110 & O3\,If$^{\ast}$/WN & d, k & O2\,If$^{\ast}$ & 12.7 & 0.2  &m & 18.45 & LMC & --6.0 & --11.2 \\
Mk 35  & BAT99-114 & O3\,If$^{\ast}$/WN & j, k & O2\,If$^{\ast}$/WN5 & 12.7     & 0.25:   &o & 18.45 & LMC&--6.0    & --11.2 \\ 
Mk 30  & BAT99-113 & O3\,If$^{\ast}$/WN  & j, k& O2\,If$^{\ast}$/WN5 & 12.8  & 0.25:   &o & 18.45 & LMC&--5.9 & --11.1\\ 
Mk 37Wb & BAT99-104& O3\,If$^{\ast}$/WN & k & (O2\,If$^{\ast}$/WN5)   & 13.06     &  0.3  &o & 18.45 & LMC& --5.7  & --10.9 \\ 
Cyg OB2--22A&        & O3\,If$^{\ast}$     & h &     & 6.2 & 0.6: & p&11.3 & Cyg OB2&--5.7 & --10.4 \\
TSWR3 & BAT99-93  & O3\,If$^{\ast}$/WN6 & j & O3\,If$^{\ast}$ & 13.35 & 0.1:& & 18.45 & LMC & --5.2 & --9.9 \\
Mk 51  & BAT99-97  & O3\,If$^{\ast}$/WN7 & j & O3.5 If$^{\ast}$/WN7 & 13.77 & 0.25:  &o & 18.45 & LMC&--4.9 & --8.6  \\ 
HD 193077 & WR138 & WN5o & a & WN6o & 6.6 & 0.1 && 11.3 &Cyg OB1?& --4.8 & --8.9  \\
Sk --67$^{\circ}$ 22& BAT99-12 & O3\,If$^{\ast}$/WN6, O2\,If$^{\ast}$ &l, b  & O2\,If$^{\ast}$/WN5 & 13.8 & 0.05 
& &18.45 & LMC&--4.7 & --9.9 \\
\hline
\end{tabular}
\end{center}
(a) Smith et al. (1996); (b) Massey et al. (2005); (c) Schnurr et al. 
(2008a); (d) Crowther \& Dessart (1998); (e) Schnurr et al.  (2008b); (f) 
Shara et al. (1991); (g) Rauw et al. (2004); (h) Walborn et al. (2002); 
(i) Ma\'{i}z
Apell\'{a}niz et al. (2007); (j) Walborn \& Blades (1997); (k) 
Massey \& Hunter (1998); (l) Walborn (1982a); (m) Crowther et al. (2010); (n) Tapia 
et al. (1988); (o) Campbell et al. (2010); (p) Torres-Dodgen et al. (1991)
\end{table*}

\section{Near-IR spectroscopy of Of/WN stars}

Classification of early-type stars has historically relied upon high 
quality blue visual spectroscopy, to which UV morphological sequences have 
been added (e.g. Walborn et al. 1992). More recently, the advent of 
efficient detectors and large ground-based telescopes has opened up the 
near-IR window (primarily K-band) for spectral typing, albeit generally 
cruder with respect to optical spectroscopy (Gray \& 
Corbally 2009). 

This is especially relevant for 
emission line early-type stars, which are readily discovered either from 
near-IR  narrow-band surveys (Crowther et al. 2006; Shara et al. 2009) or 
near to mid-IR spectral energy distributions (Hadfield et al. 2007). In 
addition, 
spectroscopically identifying individual stars within crowded fields from 
the ground -- such as dense, star clusters -- favours Adaptive Optics 
which  is significantly more effective in the near-IR than at  optical 
wavelengths (e.g. Schnurr  et al. 2008b). Consequently, can one 
distinguish between Of, Of/WN and WN stars solely from near-IR 
spectroscopy?

\begin{figure*}
\begin{center}
\includegraphics[width=0.9\textwidth,clip]{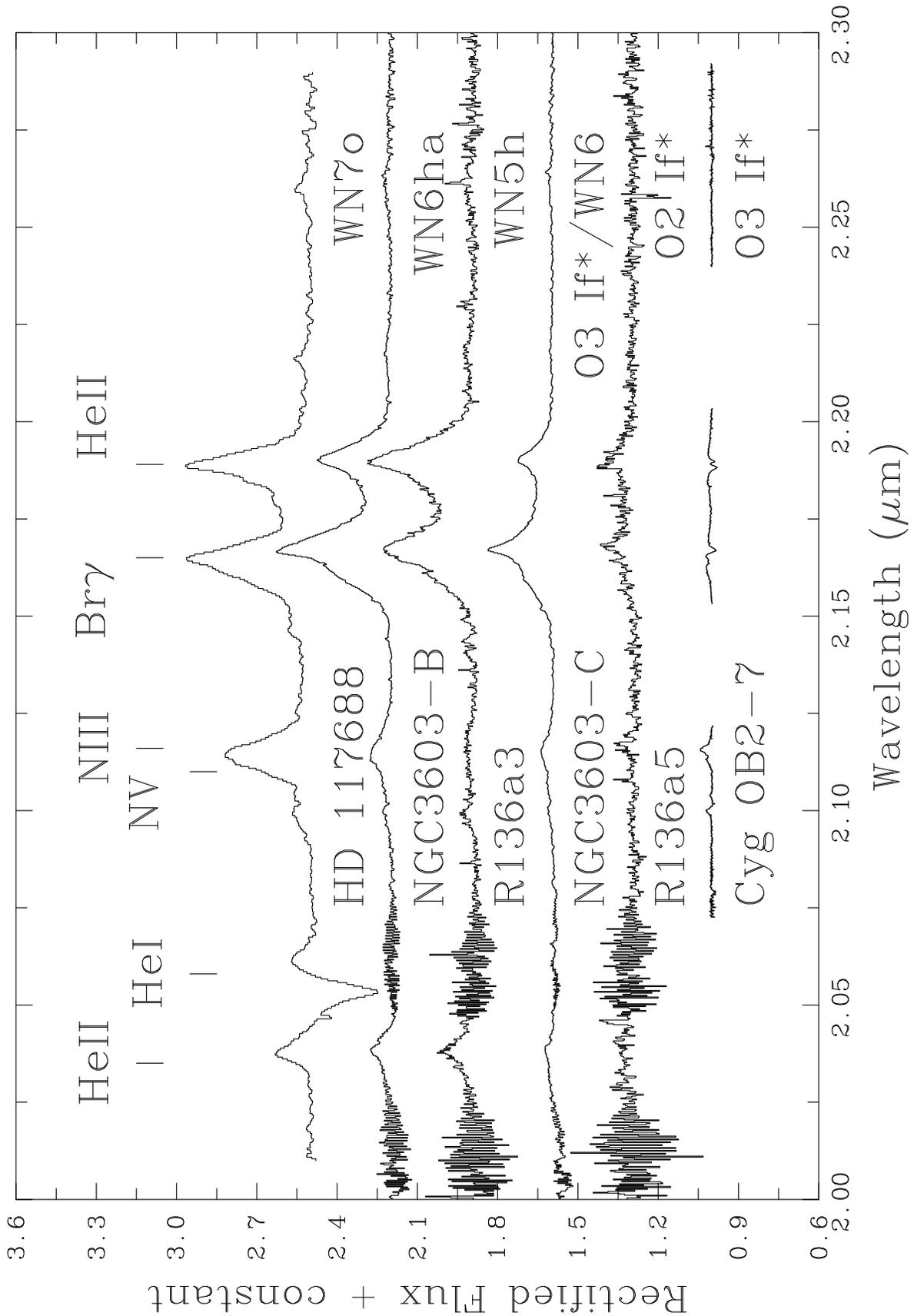}
\caption{Rectified, K-band  spectrograms of early-type O supergiants, Of/WN stars and WN5--7 
stars. Stars are uniformly offset by 0.5 continuum units for clarity.}
\label{k_band}
\end{center}
\end{figure*}

We present a montage of selected Of, Of/WN and WN5--7 stars in 
Figure~\ref{k_band}, drawn from Hanson et al. (2005), Schnurr et al. 
(2008b, 2009) plus unpublished NTT/SOFI spectroscopy of HD 117688 (WR55, 
WN7o) from N. Homeier (priv. comm). From Smith et al. (1996), the 'o'
indicates the absence of hydrogen from late-type WN stars on the basis
of the Pickering-Balmer decrement. It is apparent that the Of and WN5--7 
stars possess, respectively, the weakest and strongest Br$\gamma$ and 
He\,{\sc ii}  2.189$\mu$m emission, as anticipated. Emission features 
from NGC 3603-C (O3\,If$^{\ast}$/WN6), the sole transition star for which we 
possess high quality K-band spectroscopy, are intermediate between these 
extremes. 

Hanson et al. (1996, 2005) note that detailed classification of  early O
supergiants is  not possible solely from K-band spectroscopy. Indeed, 
C\,{\sc iv}  2.069/2.078$\mu$m emission is seen in some (e.g. HD 93129A, 
O2\,If$^{\ast}$; HD  15570, O4\,If) but not all cases (Cyg OB2--7, O3\,If$^{\ast}$). In 
contrast, N\,{\sc v}  2.110$\mu$m/N\,{\sc iii} 2.116$\mu$m serves as a 
primary  diagnostic for  weak-lined WN4--6 stars at near-IR wavelengths 
(Crowther  et al. 2006),  since He\,{\sc ii} 2.189$\mu$m/Br$\gamma$ is 
strongly modified by hydrogen  content within this spectral range. 
He\,{\sc i} 2.058$\mu$m is generally observed as a P Cygni profile for 
subtypes later than WN6, although this is absent for weak-lined WN stars
(e.g. NGC 3603-B WN6h).

Divisions between Of, Of/WN and WN5--7 stars from near-IR spectroscopy are
less definitive than from visual diagnostics, in view of the relatively 
small sample of stars for which high quality datasets are available. 
Nevertheless, the sum of the equivalent widths of
Br$\gamma$ + He\,{\sc ii} 2.189$\mu$m emission lines for 
all WN5--7 stars for which optical and near-IR spectroscopy 
is available is in excess of $\sim$60\AA. On contrast, the sum of 
Br$\gamma$ + He\,{\sc ii} 2.189$\mu$m lies in the range $W_{\lambda}$ = 
2--20\AA\ for typical  early Of supergiants and $\sim$40\AA\ for 
NGC\,3603-C (O3\,If$^{\ast}$/WN6). 

Regarding approximate boundaries between subtypes analogous to those 
presented for He\,{\sc ii} $\lambda$4686 in Fig.~\ref{4686}, it is likely 
that this occurs close to W$_{\lambda}$(Br$\gamma$+He\,{\sc ii} 
2.189$\mu$m) $\sim$30\AA\ for the transition between Of and 
Of/WN5--7 stars, and  W$_{\lambda}$(Br$\gamma$+He\,{\sc ii} 2.189$\mu$m) $\sim$50\AA\ 
for the boundary between Of/WN and WN5--7 stars. R136a5 (O2\,If$^{\ast}$) 
and NGC\,3603-C (O3\,If$^{\ast}$/WN6) would then lie relatively close to 
these boundaries, such that ambiguous classification could result solely from 
near-IR  spectroscopy. However, such thresholds would certainly 
support a WN7  Wolf-Rayet subtype for W43 \#1 as proposed by 
Blum et al. (1999), for 
which EW(Br$\gamma$+He\,{\sc ii} 2.189$\mu$m) $\sim$65\AA\ as measured
from our own unpublished VLT/ISAAC spectroscopy.

Very late-type WN stars complicate the picture at near-IR wavelengths, as in
the optical. Such stars possess weak (typically P Cygni) He\,{\sc ii} 
2.189$\mu$m emission, plus narrow relatively weak Br$\gamma$ emission (e.g.
Bohannan \& Crowther (1999) measured EW(Br$\gamma$)$\sim$20\AA\ 
for WN9ha stars).  This  morphology is common to some early-type 
Of  supergiants, such as HD  16691 (O4\,If, Conti et al. 1995), albeit 
with W$_{\lambda}$(Br$\gamma$) $\sim$7\AA. O\,If$^{\ast}$/WN5--7 stars can be 
discriminated from such stars through the  simultaneous presence of 
prominent emission at Br$\gamma$ {\it and} He\,{\sc ii} 2.189$\mu$m, with 
intermediate Br$\gamma$ equivalent widths. In addition, some WN8--9 stars 
exhibit P Cygni profiles at He\,{\sc i} 2.058$\mu$m, although this is 
extremely weak for WN9ha stars. 

This discussion is relevant to early-type emission line stars within 
visually obscured clusters, such as the Arches (Figer et al. 2002). From an 
assessment of K-band  spectroscopic datasets for the Arches stars 
presented by Martins et al. (2008) there are no examples of 
O2--3.5\,If$^{\ast}$/WN5--7 stars in the Arches cluster, based on our criteria set 
out here.

\section{Evolutionary status of Of/WN stars}\label{discussion}

Table~\ref{catalog} provides photometric properties of selected Of, Of/WN 
and WN stars, sorted by absolute K-band magnitude.  We prefer to rank stars 
by  absolute K-band magnitude instead of the more usual V-band, due to their 
reduced extinction corrections. We are also able to provide 
{\it qualitative} estimates of stellar luminosities using a calibration of
K-band bolometric  corrections, BC$_{K}$, presented in  Table~\ref{bc}.
These are based on spectroscopic results obtained with the non-LTE CMFGEN 
code (Hillier \& Miller  1998) for NGC 3603-C (O3\,If$^{\ast}$/WN6), 
R136a2 (WN5h) and NGC 3603-A1 (WN6h) from Crowther et al. (2010), 
Melnick 42 (O2\,If$^{\ast}$) and Sk --67$^{\circ}$ 22 (O2\,If$^{\ast}$/WN5)
from Doran \& Crowther (2011) plus O3--4 supergiants 
from Martins \& Plez (2006).

If we assume that the estimated bolometric correction 
for Sk --67$^{\circ}$ 22 is representative of 
O2\,If$^{\ast}$/WN5 stars, this group will typically possess high luminosities,
e.g. $M_{\rm Bol} \sim -11.2$ mag or $\log L/L_{\odot} \sim 6.4$ for Melnick
35. Based upon the main-sequence evolutionary models presented in Crowther
et al. (2010), the properties of most O2\,If$^{\ast}$/WN5 stars are consistent with
very massive ($M_{\rm init}$ $\sim$ 150 $\pm$ 30 $M_{\odot}$), rotating
stars at a relatively small age of $\sim$1 Myr (e.g. Fig.~1, 
Doran \& Crowther 2011). Such stars rapidly develop powerful stellar 
winds at a very early phase in their evolution due to their proximity to 
the Eddington limit, such that they {\it may} resemble O2 giants (e.g.  
HDE\,269810, O2\,III(f*)) at the zero-age main sequence, transitioning
through the  Of/WN stage before entering the hydrogen-rich WN phase 
(Crowther et al.  2010) while still in a core hydrogen-burning phase. 
Recall from Walborn et 
al. (2002) that O2 dwarfs typically possess masses substantially inferior 
to 100  $M_{\odot}$, while some Of/WN stars are members of very high mass
binary systems (e.g. WR20a, Rauw et al. 2004, 2005).

However, not all Of/WN stars are exceptionally massive, young stars. 
From Table~\ref{catalog}, the properties estimated for Sk --67$^{\circ}$ 22 
(O2\,If$^{\ast}$/WN5) 
by Doran \& Crowther (2011) reveal a much lower luminosity of
$M_{\rm Bol} \sim -9.9$ mag or $\log L/L_{\odot} \sim 5.9$.
In  contrast with  the high luminosity/high mass majority, such Of/WN stars 
are presumably  the immediate precursors of classical hydrogen-deficient 
WN stars, and already at an relatively advanced evolutionary phase, with lower
initial masses ($\sim$60 M$_{\odot}$) and somewhat older ages ($\geq$ 2.5 Myr). 

Adopting a K-band bolometric correction of $-3.7$ mag, 
Mk 51 (O3.5\,If$^{\ast}$/WN7) 
would have a yet lower luminosity of $M_{\rm Bol}$ = --8.6 mag or 
$\log L/L_{\odot} \sim 5.3$. Presumably Mk 51 has either 
evolved through a red supergiant or Luminous Blue Variable phase prior to 
returning to the blue part of the Hertzsprung-Russell diagram, or 
such a low luminosity supergiant might be a post-mass transfer binary
(Walborn et al. 2002).

Morphologically we are unable to discriminate between the high and low 
luminosity Of/WN stars. More quantitative results await the detailed 
analysis of such stars which  is presently underway within the context of 
the VLT-FLAMES Tarantula Survey (J.M.~Bestenlehner et al., in 
preparation).

From Table~\ref{catalog}, it is apparent that the LMC hosts the 
majority of transition stars. We do not anticipate a substantial difference 
between the  wind or physical properties  of LMC early type stars with respect to the Galaxy 
as a  result of the factor of $\sim$2 reduced metallicity, $Z$. Radiatively driven wind 
theory  (Vink et al. 2001) and observations (Mokiem et al. 2007) suggest a mass-loss scaling 
$\propto Z^{\sim0.7}$, such that LMC stars would be expected to possess 
slightly weaker stellar 
winds than their Milky Way counterparts. Therefore, the LMC
incidence of transition stars with respect to bona fide WN stars is 
anticipated to be modestly higher than the Milky Way. 

In fact, taking our revisions into account, O2--3.5\,If$^{\ast}$/WN5--7 
comprise 7\% of the $\sim$106 WN-flavoured LMC Wolf-Rayet stars listed by 
Breysacher et al. (1999).\footnote{Ten stars were listed as `hot' 
transition stars by Breysacher et al. (1999) to which we add Melnick\,37a
and AB4, with TSWR3, Melnick 42 and R136a5 removed.  In addition, BAT99-80 
(TSWR2, NGC 2044W-9A), originally assigned O4\,If/WN6 (Testor \& Schild 
1990) has subsequently been revised to O4\,If$^{+}$ (Walborn et al. 1999). 
We also include the newly discovered WN star VFTS-682 from Evans et al. 
(2011) in our LMC statistics.} In contrast, transition stars (SMSP\,2, 
HD\,93162, HD\,97950-C) comprise only 2\% of the (highly incomplete) 
$\sim$175  WN stars compiled for the Milky Way by van der Hucht (2001, 
2006).

Reduced wind densities are only expected to partially explain these 
differences. If Of/WN subtypes arise preferentially in very massive stars, 
one may expect an excess of transition stars in the most massive, young 
clusters. Indeed, the region of 30 Doradus close to R136 dominates
the known Of/WN population, since within the Milky Way relatively modest 
star-forming regions such as Carina and NGC\,3603 are accessible to 
optical spectroscopic surveys.

\begin{table}
\begin{center}
\caption{Calibration of K-band bolometric corrections for early-O and WN5--7 stars
based upon CMFGEN model atmosphere analyses.}
\label{bc}
\begin{tabular}{l@{\hspace{1.5mm}}l@{\hspace{1.5mm}}l@{\hspace{1.5mm}}l@{\hspace{1.5mm}}l@{\hspace{1.5mm}}
l@{\hspace{1.5mm}}l@{\hspace{1.5mm}}l@{\hspace{1.5mm}}l}
\hline
Star & Subtype & $T_{\rm eff}$ & BC$_{K}$ & Ref \\
      &        & kK          & mag & \\
\hline
HD 92740     & WN7ha   & 38 & --3.7: & unpublished \\
           & O4\,I     & 40 & --4.55 & Martins \& Plez (2006) \\
NGC 3603-A1b & WN6h   & 40 & --4.1 & Crowther et al. (2010) \\
             & O3\,I   & 42 & --4.69 & Martins \& Plez (2006) \\
NGC 3603-C &O3\,If$^{\ast}$/WN6& 44 & --4.4 & Crowther et al. (2010) \\
Mk 42 & O2\,If$^{\ast}$        & 50 & --5.2 & Doran \& Crowther (2011) \\
Sk --67$^{\circ}$ 22 & O2\,If$^{\ast}$/WN5 & 49 & --5.2 & Doran \& Crowther (2011) \\
R136a2 & WN5h         & 53 & --4.9 & Crowther et al. (2010) \\
\hline
\end{tabular}
\end{center}
\end{table}

\begin{figure*}
\begin{center}
\includegraphics[width=0.9\textwidth,clip]{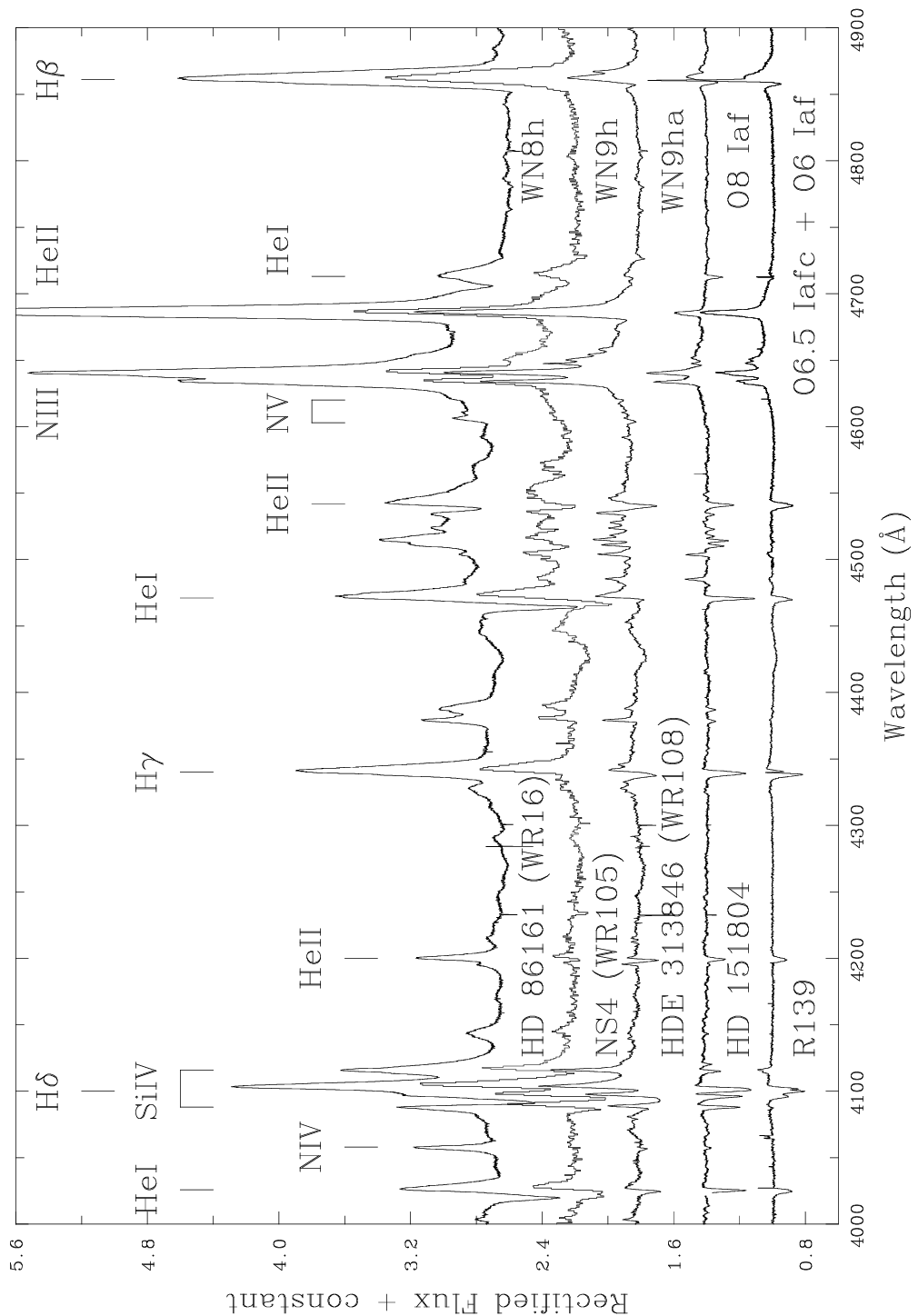}
\caption{Rectified, blue-violet spectrograms of stars spanning O6.5--8\,If through 
WN8--9. Stars are uniformly offset by 0.3 continuum units for clarity.}
\label{wn8-9}
\end{center}
\end{figure*}

\section{Lower ionization spectra (N\,{\sc iv} $\lambda$4058 
emission weak/absent)}\label{oddities}

We have sought to discriminate O2--3.5\,If$^{\ast}$/WN5--7 from 
O2--3.5\,If$^{\ast}$ and WN5--7 stars through the presence of H$\beta$ 
emission, providing N\,{\sc iv} $\lambda$4058 is present in emission. 
However,   for completeness the potential implications of our criteria 
for stars in which N\,{\sc iv} $\lambda$4058 is weak/absent also need to 
be considered, 
including a second class of star historically classified as Of/WN.


This second flavour of `/' star, was introduced by Walborn (1982b) 
and Bohannan \& Walborn (1989) to refer to another category of 
peculiar stars, assigned Ofpe/WN9. In contrast with the {\it intermediate} 
O2--3.5\,If$^{\ast}$/WN  subtypes, the
Ofpe/WN9 classification was intended to denote {\it alternative} 
descriptions/interpretations for the same object. Indeed, Walborn (1977)
had earlier suggested {\it either} a O\,Iafpe or WN9 (or WN10) classification 
for one such star, HDE 269227. The latter designation was preferred 
by Smith et al. (1994) who proposed WN9--11 to distinguish between stars
of varying ionization, while Bohannan \& Crowther (1999) also 
argued that Ofpe stars should be reclassified as WN9ha. 
Nevertheless, Ofpe/WN9 remains in common usage both for surveys of 
external galaxies (e.g.  Bresolin et al. 2002) and highly reddened stars 
within the inner Milky Way (e.g. Mason et al. 2009).


A spectral montage of late-type Of and WN8--9 stars is displayed in 
Fig.~\ref{wn8-9} (see also Chapter 3 of Gray \& Corbally 2009). The 
spectral morphology of mid to late-type Of stars resembles late 
WN stars in the vicinity of He\,{\sc ii} $\lambda$4686. We also note that
P Cygni H$\beta$ {\it is} relatively common in late Of  supergiants, 
including R139 (O6.5\,Iafc + O6\,Iaf, Taylor  et al. 2011),  HD\,151804 
(O8\,Iaf, Crowther \& Bohannan 1997) and He\,3--759 (O8\,If, Crowther \& 
Evans 2009).

Analogously to O2--3.5\,If$^{\ast}$/WN transition stars, we have 
considered the possibility of an  intermediate category for 
stars in which N\,{\sc iv} $\lambda$4058  emission is weak/absent. Recall 
that Wolf-Rayet spectral types 
are intended for predominantly emission line stars at visible wavelengths, 
while O spectral types  are  appropriate for primarily absorption line 
stars. In contrast with `hot' transition stars, late-type Of and WN 
stars can be cleanly distinguished in  Fig.~\ref{wn8-9}.
Specifically, WN8--9 stars exhibit  strong P Cygni  
He\,{\sc i}  $\lambda$4471, versus absorption in late-type  Of stars. 
Walborn  (1975)  has previously  highlighted the  development of  P Cygni 
He\,{\sc i} $\lambda$5876 from  HD 151804  (O8\,Iaf)  and HD 
152408 (O8\,Iafpe or WN9ha) to HD  151932 (WN7h).  Other 
morphological differences include He\,{\sc ii}  $\lambda$4542, 
$\lambda$4200 and the complex around H$\delta$.

On the basis of  presently available observations, we therefore
propose restricting {\it intermediate} Of/WN classifications solely to
the earliest O subtypes. For lower ionization stars in which N\,{\sc iv} 
$\lambda$4058 is weak/absent, yet each of He\,{\sc ii} $\lambda$4686, 
N\,{\sc iii} $\lambda\lambda$4634-41  and H$\beta$ are in emission, we 
favour:
\begin{enumerate} 
\item Adhering to existing WN subtypes if the morphology of He\,{\sc 
i}  $\lambda$4471 is a P Cygni profile (e.g. NS4, WN9h) or WNha if, 
in addition, the morphology of He\,{\sc ii} $\lambda$4541, $\lambda$4200 are 
P Cygni profiles (e.g. HDE\,313846, WN9ha).
\item Retaining existing O  supergiant spectral types if the 
morphology of  He\,{\sc i} $\lambda$4471  is in absorption (e.g. 
HD\,151804,  O8\,Iaf)\footnote{An O4\,Iaf subtype is retained
for R136b (Massey \& Hunter 1998) since He\,{\sc i} $\lambda$4471 is 
observed in absorption in HST/FOS spectroscopy, in spite of H$\beta$  
emission. Crowther \& Dessart (1998) had tentatively proposed WN9ha for 
R136b,  although an earlier WN8 subtype would have been more appropriate 
since N\,{\sc iv} $\lambda$4058 is detected at a 4--5$\sigma$ level.}.
\end{enumerate} 


\section{Summary}

We present a revised classification scheme for O\,If$^{\ast}$/WN5--7 
stars, in order to take into account various revisions to the Of$^{\ast}$ 
(Walborn et al. 2002) and mid-WN (Smith et al. 1996) subtypes since the 
initial introduction of this subclass (Walborn 1982a).

\begin{enumerate}

\item We propose that O2--3.5\,If$^{\ast}$, O2--3.5\,If$^{\ast}$/WN5--7 
and WN5--7 stars may be discriminated using the morphology of H$\beta$: 
purely in absorption for O2--3.5\,If$^{\ast}$ stars;  P Cygni for 
O2--3.5\,If$^{\ast}$/WN5--7 stars; purely in emission for WN stars.

\item Based upon our updated scheme at least ten Of/WN objects are 
identified in the LMC (primarily 30 Doradus) and Milky Way (Carina Nebula, 
NGC 3603, Westerlund 2).
 
\item Since many young high mass stars in the Milky Way are visually obscured due 
to dust extinction we also discuss  approximate criteria from which
early Of,  Of/WN5--7 and WN5--7 subtypes may be discriminated
 from near-IR  spectroscopy. We emphasise that  high quality blue-visual spectroscopy provides superior 
diagnostics. 

\item We suggest that the majority of O2--3.5\,If$^{\ast}$/WN5--7 stars 
are young, very massive hydrogen-burning stars, genuinely intermediate 
between O2--3.5\,If$^{\ast}$ and WN5--7 subtypes, although a minority are 
apparently lower mass core helium-burning stars evolving blueward towards 
the classical WN sequence. We suggest that transition stars form a 
larger subset of the LMC WN population than that of the Milky Way due to weaker 
stellar winds and a higher percentage of very massive stars within 30 
Doradus with respect to typical Galactic star forming regions.

\item  On the basis of presently available observations, we do 
not favour {\it intermediate} Of/WN subtypes for 
He\,{\sc ii} $\lambda$4686 emission line stars
in which N\,{\sc iii} $\lambda\lambda$4634--41 $\gg$ N\,{\sc iv} 
$\lambda$4058. We advocate: (a) WN8--9 spectral types if the morphology of 
He\,{\sc i}  $\lambda$4471 is P Cygni, or (b) mid- to late- Of 
supergiant subtypes  if He\,{\sc i} $\lambda$4471 is observed in 
absorption.

\end{enumerate}


\section*{Acknowledgements}

We are grateful to the Tarantula survey consortium for providing UVES and 
FLAMES datasets prior to publication, especially to William Taylor,
Chris Evans and Sergio Simon Diaz. 
We wish to  thank Ian Howarth for blaze correction of the HD 38282 
UVES datasets, Jesus Ma\'{i}z  Apell\'{a}niz for providing the spectrum of Pismis 
24-1NE,  Phil Massey for allowing us to inspect his observations of Brey 58/AB4, 
Nicole  Homeier for the K-band spectrum of HD 117688 and Emile Doran 
for providing selected K-band photometry. Chris Evans kindly provided 
comments on the draft manuscript  prior to submission, while we also
acknowledge important comments by the referee Ian Howarth.
Based in part on observations made with 
ESO Telescopes at the Paranal Observatory  using programme ID's 
70.D-0164, 72.C-0682, 74.D-0109, 182.D-0222. Additional observations were 
taken with the NASA/ESA Hubble Space Telescope, obtained from the data 
archive at the Space Telescope Institute. STScI is operated 
by the Association of Universities for Research in Astronomy, Inc. under 
the NASA contract NAS 5-26555.

\label{lastpage}

\end{document}